\documentclass[prd,showpacs,amsmath,amssymb,nofootinbib,superscriptaddress,notitlepage]{revtex4-1}
\usepackage[usenames,dvipsnames,svgnames,table]{xcolor} 
\usepackage{array,mathtools,amssymb,booktabs,multirow,diagbox,hhline,natbib}
\pdfoutput=1

\usepackage{graphicx,graphics,color}
\usepackage{dcolumn}
\usepackage{bm}
\usepackage{bbold}
\usepackage{hyperref}
\hypersetup{ 
    pdfnewwindow=true,      
    colorlinks=true,       
    allcolors=SeaGreen
} 
\usepackage{amsmath}
\usepackage{xcolor}
\usepackage{tikz}
\usepackage{tikz-feynman}
\usepackage{simpler-wick}
\usepackage{subfigure}

\allowdisplaybreaks

\newcommand{\Od}{{\cal O}}
\newcommand{\im}{\mbox{Im}\,}
\newcommand{\re}{\mbox{Re}\,}

\newcommand{\mean}[1]{\left\langle{#1}\right\rangle}
\newcommand{\condl}{\mean{\bar q q}_l}

\newcommand{\Sr}{\textrm{\bf S}}
\newcommand{\Tr}{\textrm{\bf T}}
\newcommand{\Ur}{\textrm{\bf U}}
\newcommand{\gsim}{\raise.3ex\hbox{$>$\kern-.75em\lower1ex\hbox{$\sim$}}}

\begin{document}

\title{Pion scattering, light resonances and chiral symmetry restoration at nonzero chiral imbalance and temperature}
\author{Angel G\'omez Nicola}
\email{gomez@ucm.es}
\affiliation{Universidad Complutense de Madrid, Facultad de Ciencias F\'isicas, Departamento de F\'isica Te\'orica and
IPARCOS, Plaza de las Ciencias 1, 28040 Madrid, Spain}
\author{Patricia Roa-Bravo}
\email{patriroa@ucm.es}
\affiliation{Universidad Complutense de Madrid, Facultad de Ciencias F\'isicas, Departamento de F\'isica Te\'orica, Plaza de las Ciencias 1, 28040 Madrid, Spain}
\author{Andrea Vioque-Rodr\'iguez}
\email{avioque@ucm.es}
\affiliation{Universidad Complutense de Madrid, Facultad de Ciencias F\'isicas, Departamento de F\'isica Te\'orica and
IPARCOS, Plaza de las Ciencias 1, 28040 Madrid, Spain}

\begin{abstract}
 We calculate the pion scattering amplitude at nonzero temperature and nonzero $\mu_5$, the chemical potential associated to chiral imbalance in a locally $P$-breaking scenario. The amplitude is calculated up to next to leading order in Chiral Perturbation Theory and is unitarized with the Inverse Amplitude Method to generate the poles of the $f_0(500)$ and $\rho (770)$ resonances. Within the saturation approach,  the thermal $f_0(500)$ pole allows to determine $T_c(\mu_5)$, the transition temperature for chiral symmetry restoration. Our results confirm the growing behaviour of $T_c(\mu_5)$ found in previous works and, through a fit to lattice results, we improve the uncertainty range of the low-energy constants associated to $\mu_5$ corrections in the chiral lagrangian. The results for the $\rho (770)$ pole are compatible with previous works regarding the dilepton yield in heavy-ion collisions.  
 \end{abstract}

 \maketitle

\section{Introduction}

The possible generation of a phase characterized by Local Parity Breaking (LPB) in nuclear matter under extreme conditions, in particular in high-energy heavy ion collisions (HICs), has been the subject of recent studies  within the so-called Chiral Magnetic Effect (CME) \cite{Kharzeev:2007jp,Fukushima:2009ft,Kharzeev:2013ffa} 
This LPB phenomenon can be attributed to the difference between the number densities of right- and left-handed chiral fermions, the so-called chiral imbalance, which has motivated  the study of effective lagrangians \cite{Fukushima:2010fe,Chernodub:2011fr, Andrianov:2012dj,Andrianov:2013dta,Yu:2015hym,Braguta:2016aov,Ruggieri:2016ejz,Andrianov:2017meh,Andrianov:2019fwz,Espriu:2020dge} and lattice QCD \cite{Yamamoto:2011ks,Braguta:2015zta,Braguta:2015owi,Feng:2017dom,Astrakhantsev:2019wnp} 
with  a nonzero chemical potential parametrizing chiral imbalance.

The non-abelian nature of the strong interaction gives rise to a complicated vacuum. As a result, the vacuum state can accommodate multiple topological sectors that are separated by high energy barriers. In the presence of a hot medium, sphaleron transitions can connect these configurations through quantum fluctuations of the vacuum state \cite{McLerran:1990de}.  Actually, a topological charge may arise in the fireball as a result of a HIC. Such charge can be defined as the space integral of the Chern-Simons current, as follows
\begin{equation}
\begin{split}
    T_5&=\int_{vol}d^3\vec{x}\,J^{CS}_0(x),\\
    J^{CS}_0(x)&=\dfrac{g^2}{32\pi^2}\epsilon_{\nu\rho\sigma}\text{Tr}\left(G^\nu\partial^\rho G^\sigma-i\dfrac{2}{3}G^\nu G^\rho G^\sigma\right)
\end{split}
\end{equation}
where the integration is over a given region within the fireball volume.

Thus, introducing into the QCD Lagrangian a chemical potential $\mu_{CS}$ in a gauge invariant way, i.e. adding $\Delta\mathcal{L}=\mu_{CS}\Delta T_5$, with
\begin{equation}
	\Delta T_5=T_5(t_f)-T_5(0)=\dfrac{g^2}{32\pi^2}\int_0^{t_f}dt\,\int_{vol}d^3\vec{x}\, G_{\mu\nu}^a\tilde{G}^{a\mu\nu}
\end{equation}
it would be possible to trigger the value of $\left\langle\Delta T_5\right\rangle$.

We can associate a non-zero topological charge with a non-trivial quark axial charge $Q_5$ integrating the $U(1)$ anomaly relation over a finite space volume
\begin{equation}
\begin{split}
	\dfrac{d}{dt}\left(Q_5-2N_f T_5\right)&=2 i \int_{vol}d^3\vec{x}\,\bar{q}\mathcal{M}\gamma_5 q\\
	Q_5&=\int_{vol}d^3\vec{x} \,\bar{q}\gamma_0\gamma_5q
\end{split}
\label{Q5}
\end{equation}
Thus, using the well-known Atiyah-Singer index theorem, which establishes a relationship between the topological charge of the gauge field and the right $N_R$ and left $N_L$ zero-eigenstates of the Dirac operator we get
\begin{equation}
    2N_f T_5= N_L-N_R
\end{equation}

For $u$ and $d$ quarks, the characteristic time of left and right quark oscillations is of  order $1/m_q$ \cite{Andrianov:2017meh}, i.e.,  significantly larger than the typical duration of the fireball. This observation suggests that, during the lifetime of the fireball, the chiral charge $Q_5$ associated with the light $u$ and $d$ quarks may remain approximately constant in a typical heavy-ion collision, since the oscillation can be neglected.

Therefore, for light $u,d$ quarks for which both the $L-R$ oscillations and the mass terms in \eqref{Q5}  are negligible \cite{Andrianov:2017meh}, one can assign, as customary, a constant axial chemical potential $\mu_5$ in order to parameterize a source of parity breaking or chiral imbalance, which couples in the QCD lagrangian to the timelike component of the $U(1)_A$ abelian axial current, i.e,

\begin{equation}
    \mathcal{L}_{QCD} \rightarrow \mathcal{L}_{QCD}+\mu_{5}\bar{q}\gamma_{5}\gamma^{0}q
    \label{Lqcdmu5}
\end{equation}
where  $\mu_5=\mu_{CS}/(2N_f)$.

Temperature effects with $\mu_5\neq 0$ introduce interesting features. The Physics behind this is pertinent since, as explained above, one expects these LPB regions to be formed within a heavy-ion collision environment where medium effects such as temperature and baryon chemical potential play a crucial role and might affect the very same stability of such regions depending on the sector of the QCD phase diagram covered  during the fireball evolution. 
 Conversely, one may ask about the effect that LPB, parametrized by $\mu_5$, may have on the phase diagram and observables from HIC.  In particular, the $\mu_5$-dependence of the QCD transition of deconfinement and chiral symmetry restoration. The latter has been actually the subject of recent lattice studies  \cite{Braguta:2015zta,Braguta:2015owi} which show a slowly increasing behaviour of the transition temperature $T_c(\mu_5)$, consistently with the NJL-model \cite{Braguta:2016aov,Ruggieri:2016ejz}
 and Chiral Perturbation Theory (ChPT) \cite{Espriu:2020dge} effective theory analyses, although in apparent contradiction with the decreasing behaviour found in previous works \cite{Fukushima:2010fe,Chernodub:2011fr}, which may come from the NJL regularization procedure \cite{Yu:2015hym}.
 An interesting suggestion in this context has been to relate  $\mu_5$ effects to the enhancement of the dilepton production rate in HIC in the low invariant mass region, close to the $\rho(770)$ mass, where $\mu_5\neq 0$ would contribute to the observed production  enhancement 
 \cite{Andrianov:2012hq,Andrianov:2014uoa,Chaudhuri:2022rwo}.

 In the recent work \cite{Espriu:2020dge},  the most general low-energy effective lagrangian at $\mu_5\neq 0$ has been derived within the Chiral Perturbation Theory (ChPT) framework \cite{Weinberg:1978kz,Gasser:1983yg} including finite $T$ effects. The energy density was derived up to Next to Next to Leading Order (NNLO), as well as  relevant quantities derived from it such as the quark condensate signaling chiral symmetry breaking, the chiral density and the topological susceptibility. One of the conclusions of that work is that new terms appear in the lagrangian which therefore generate new Low-Energy Constants (LEC). The numerical value of those constants were fixed in \cite{Espriu:2020dge} to the lattice results for $T_c(\mu_5)$, using the quark condensate, the topological susceptibility and the chiral density. 

 Here we will extend the analysis in \cite{Espriu:2020dge} by calculating the pion-pion elastic scattering amplitude with the   $\mu_5\neq 0$ lagrangian, up to NLO, i.e. $\Od(p^4)$, within the ChPT framework and unitarizing it in order to obtain the lightest resonant states $f_0(500)/\sigma$ and $\rho (770)$. The NLO ChPT amplitude at $T=\mu_5=0$ was first derived in \cite{Gasser:1983yg} and its extension to nonzero temperature was obtained in \cite{GomezNicola:2002tn}. Here we will derive the NLO amplitude at nonzero $T$ and $\mu_5$. That ChPT amplitude will be unitarized through the Inverse Amplitude Method (IAM) \cite{Dobado:1996ps,GomezNicola:2001as} which will allow us to study the combined dependence with $T$ and $\mu_5$ of the light resonances $f_0(500)$ and $\rho (770)$ poles.  In particular, from the $f_0(500)$ thermal pole, following a resonance saturation approach for the scalar susceptibility \cite{Nicola:2013vma,Ferreres-Sole:2018djq}, one of the main signals of chiral symmetry restoration, we will obtain the transition temperature $T_c(\mu_5)$, which will allow us to  pin down the new LEC by comparison with lattice predictions and to test the robustness of previous theoretical analyses. On the other hand, the results for the $\mu_5$ dependence of the $\rho(770)$ pole will be useful to test the results  about   LPB in the dilepton spectrum  \cite{Andrianov:2012hq,Andrianov:2014uoa,Chaudhuri:2022rwo}.

The paper is organized as follows. In section \ref{sec:scatt} we calculate the $\mu_5$ corrections to the $\pi\pi$ elastic scattering amplitude within the ChPT framework including also  its temperature dependence and partial wave unitarization within the IAM. The saturated approach allowing to obtain the scalar susceptibility, and hence $T_c(\mu_5)$, from the $f_0(500)$ pole, will be discussed in section \ref{sec:Tcmu5}.  In section \ref{sec:num}  we present our numerical results. First, we will discuss how to combine our present approach  with  previous works in order  to fit  $T_c(\mu_5)$ to the lattice values and improve the new LEC determination. Second, we will provide  numerical results for the $\mu_5$ dependence of phase shifts and resonance parameters. Our conclusions are summarized in section \ref{sec:conc}.

\section{The ChPT $\pi\pi$ scattering amplitude at nonzero $T$ and $\mu_5$}
\label{sec:scatt}

To start with, we consider the most general ChPT meson low-energy lagrangian $\mathcal{L}_{eff}=\mathcal{L}_{2}+\mathcal{L}_{4}+\cdots$, taking into account local parity violating terms parametrized by an axial chemical potential or chiral imbalance $\mu_{5}$, up to $\mathcal{O}(p^4)$ order in the chiral expansion for two light flavours, derived in  \cite{Espriu:2020dge}.

The $\mathcal{O}(p^2)$ order lagrangian $\mathcal{L}_2$ does not depend on the chemical potential $\mu_{5}$, except for a constant term irrelevant for our purposes here, so we are interested in the next order $\mathcal{O}(p^4)$ lagrangian, given by

\begin{eqnarray}
    \mathcal{L}_{4}(\mu_{5}) = \mathcal{L}_{4}(\mu_{5}=0) + \kappa_{1}\mu_{5}^{2}tr(\partial^{\mu}U^{\dagger}\partial_{\mu}U) + \kappa_{2}\mu_{5}^{2}tr(\partial_{0}U^{\dagger}\partial^{0}U) + \kappa_{3}\mu_{5}^{2}tr(\chi^{\dagger}U+U^{\dagger}\chi)+ \kappa_{4}\mu_{5}^{4}
    \label{L4mu5}
\end{eqnarray}
where $\chi=2B_{0}\mathcal{M}$ with $\mathcal{M}$ the quark mass matrix $\mathcal{M}=\begin{pmatrix} 
    m_{u} & 0 \\
    0 & m_{d}
\end{pmatrix}$, $U=e^{i\Pi/F}$, with $\Pi$ the pion field matrix in the charge basis,

 \begin{equation}
     \Pi = \begin{pmatrix}
     \pi^{0} & \sqrt{2}\pi^{+}\\
     \sqrt{2}\pi^{-} & -\pi^{0}
     \end{pmatrix},
 \end{equation} and  $F$, $M_0^2=B_0(m_u+m_d)$ are respectively the pion decay constant $F_\pi$ and the squared mass $M_\pi^2$ to LO in the chiral expansion.  The $\kappa_i$ are undetermined new LEC, which are  finite since all $\Od(p^4)$ loop corrections are $\mu_5$-independent.  Here, we will work in the isospin limit $m_u=m_d$.

As in  previous analyses at finite temperature 
\cite{Quack:1994vc,Kaiser:1999mt,GomezNicola:2002tn,Loewe:2008kh,GomezNicola:2023rqi}, the scattering amplitude $\mathcal{T}$ is defined by connecting the four-point Green function with the $T$-matrix element through the Lehman-Symanzik-Zimmerman (LSZ) reduction formula, which
extracts the residue of the Green function at the poles given by the dispersion relation of the external legs \cite{IZbook}. In doing so, one has to take into account the modification of the free particle dispersion relation due to self-interactions, which introduce additional $T$ and $\mu_5$ corrections as we detail below. 

As customary, we parametrize the $T$-matrix element for $\pi^a \pi^b\rightarrow \pi^c \pi^d$ scattering once isospin and crossing symmetry are taken into account, as

\begin{eqnarray}
\mathcal{T}_{abcd}=A(\Sr,\Tr,\Ur)\delta_{ab}\delta_{cd} + A(\Tr,\Sr,\Ur)\delta_{ac}\delta_{bd}
    \nonumber  + A(\Ur,\Tr,\Sr)\delta_{ad}\delta_{bc},
\end{eqnarray}
where $\Sr=P_a+P_b=(S_0,\vec{S})$, $\Tr=P_a-P_c=(T_0,\vec{T})$ and $\Ur=P_a-P_d=(U_0,\vec{U})$ are generalized Mandelstam variables while $s=\Sr^2,t=\Tr^2,u=\Ur^2$ are the usual ones. In this way, we take into account that both $\mu_5$ and temperature effects break Lorentz covariance. In the above equation, $A(\Sr,\Tr,\Ur)$ is the $T$-matrix element corresponding to  the $\pi^{+}\pi^{-}\rightarrow \pi^{0}\pi^{0}$ process, which we can express as a perturbative series in powers $A=A_{2}+A_{4}+\dots$ where $A_{n}$ is the amplitude  of $\mathcal{O}(p^n)$. As we commented in the introduction, in this work we are interested in the NLO, i.e., up to $\mathcal{O}(p^4)$.

\begin{figure}[h]
\includegraphics[width=0.8\textwidth]{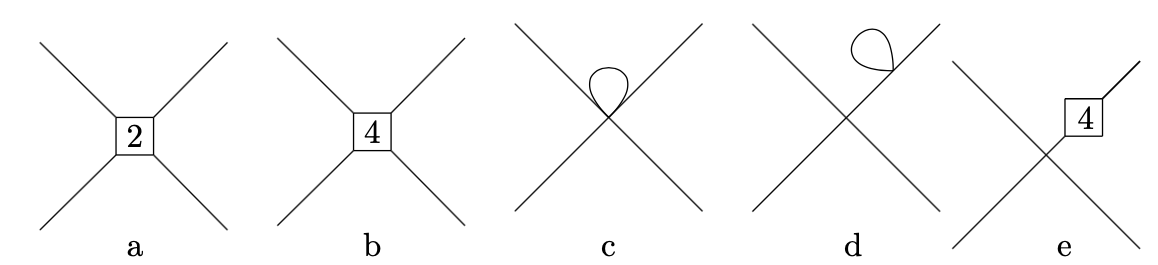}\\
\includegraphics[width=0.8\textwidth]{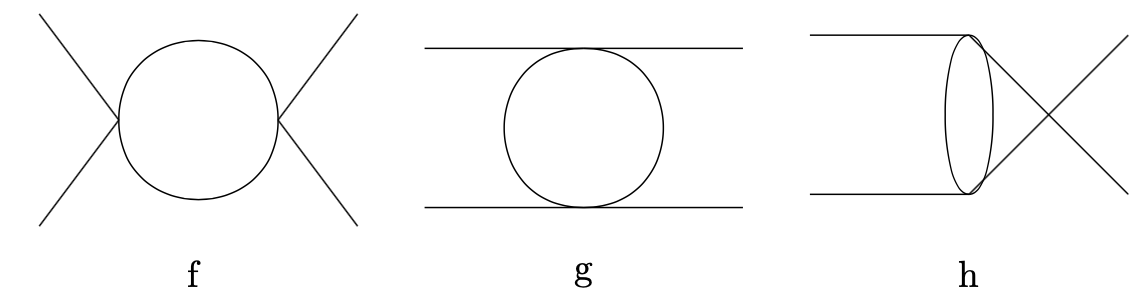}
 \caption{Diagrams contributing to pion scattering in ChPT up to $\Od(p^4)$.} 
 \label{fig:diag}
 \end{figure}

The corresponding Feynman diagrams contributing to that order are displayed in Fig.\ref{fig:diag}, where diagram a is the $\Od(p^2)$ contribution to the amplitude coming from $\mathcal{L}_2$ at tree level. The remaining diagrams provide the $\Od(p^4)$ correction. Thus, diagram b comes from $\mathcal{L}_4$ at tree level, diagram c is the one-loop tadpole correction coming from the $\Od(p^2)$ six-pion vertex, diagrams d and e account for the renormalization of the external legs at this order and therefore give rise to the modification of the dispersion relation, while diagrams f,g,h correspond, respectively, to the $S,T,U$ channels of the one-loop contributions with two $\Od(p^2)$ four-pion vertices. 

On the one hand,  the $\mu_{5}$ corrections to the amplitude come  from the following sources: 

  i) the $\mathcal{O}(p^4)$ tree level order amplitude from the four-pion terms in the lagrangian \eqref{L4mu5}, corresponding to the topology of  Fig.\ref{fig:diag}b; 
  
  ii) the correction coming from the $\mathcal{O}(p^4)$ dispersion relation modification arising from the two-pion fields in \eqref{L4mu5}, represented by diagrams d,e in Fig.\ref{fig:diag}, which only affects the LO $\mathcal{O}(p^2)$ amplitude since the corrections to the free dispersion relation are of $\Od(p^4)$

  iii) the $\mathcal{O}(p^4)$ correction coming from the residue of the LSZ formula application, as discussed below. 

  On the other hand, the corrections at finite temperature $T$ come from the loop diagrams c-h in Fig.\ref{fig:diag} and are independent of $\mu_5$ as they are the $\Od(p^2)$ vertices in those diagrams. 

We can therefore write the $T$ and $\mu_5$ dependent ChPT $\Od(p^4)$ amplitude as follows:
\begin{eqnarray}
    A (\Sr,\Tr,\Ur;T,\mu_5) &=& A_2 (s,t,u) + A_{4,tree}(s,t,u) +  A_{4,loop} (\Sr,\Tr,\Ur;T)\nonumber\\
    &+&  \delta A_{4,tree}(\Sr,\Tr,\Ur;\mu_5)+ \delta A_{4,dr}(\Sr,\Tr,\Ur;\mu_5)+ \delta A_{4,LSZ}(\Sr,\Tr,\Ur;\mu_5)
    \label{Agen}
    \end{eqnarray}
where $A_2$ is the $\Od(p^2)$ tree-level contribution corresponding to Fig.\ref{fig:diag}a. Note that the  contribution of the loops is independent of $\mu_5$ at the order considered here since the  $\mu_5$ correction to the interaction vertices starts at $\Od(p^4)$. Therefore, at this order we can import the $T$-dependent corrections to the amplitude in $A_{4,loop} (\Sr,\Tr,\Ur;T)$ directly from the calculation in \cite{GomezNicola:2002tn}.   We will calculate below the new $\mu_5$-dependent contributions $\delta A_4$ to the amplitude.

\subsection{Tree level contributions}

This contribution can be extracted directly from  the $\mathcal{O}(p^4)$  lagrangian in \eqref{L4mu5}. We get:

\begin{align}
    \delta A_{4,tree}(\Sr,\Tr,\Ur;\mu_5) = \frac{1}{3F^{2}}[3\tilde\kappa_{1}s + M_{0}^{2}(\tilde\kappa_{3} - 4\tilde\kappa_{1}) 
    + \tilde\kappa_{2}(2S_{0}^{2}-T_{0}^{2}-U_{0}^{2})]
\label{A4tree}
\end{align}
with $\displaystyle\tilde{\kappa_{i}}=\frac{4\kappa_{i}\mu_{5}^{2}}{F^{2}}$. Note that the $S_0,T_0,U_0$ terms above come from the separation of the timelike and spacelike components in the lagrangian derivatives, as a result of the loss of Lorentz covariance.

\subsection{Dispersion relation correction}

At the order we are considering, including the contributions of diagrams d,e in  Fig.\ref{fig:diag} is equivalent to consider for the external legs   the $\Od(p^4)$ renormalized pion propagator 

\begin{equation}
    G_\pi(p^2,p_0^{2}) = \frac{i Z_\pi (T)}{p^{2}-M^{2}_{\pi}(T) + \tilde{\kappa_{1}}p^{2} + \tilde{\kappa_{2}}p_0^{2} - \tilde{\kappa_{3}}M_{0}^{2}}.
\label{fullprop}
\end{equation}
where  $M_\pi^2 (T)=M_0^2\left[1+\delta_\pi (T)\right]$  and $Z_\pi=1+\delta_Z (T)$ are respectively the pion mass squared and wave function renormalization at NLO for $\mu_5=0$, the $T$-dependence coming from the tadpole diagram in Fig.\ref{fig:diag}d. Those  corrections have been included in the finite-$T$ calculation of the scattering amplitude in  \cite{GomezNicola:2002tn} and are $\mu_5$-independent. The $\tilde \kappa_i$ terms in \eqref{fullprop} come from the $\mu_5$-dependent two-pion field contributions in \eqref{L4mu5}, also included in  Fig.\ref{fig:diag}e.

The pole of \eqref{fullprop} yields the $\mu_5$-dependent modification of the pion dispersion relation at this order:

\begin{equation}
    p^{2} = M_\pi^2 (T)- \tilde{\kappa_{2}}p_{0}^{2} - (\tilde{\kappa_{1}}-\tilde{\kappa_{3}}) M_0^2
\label{disprel}
\end{equation}
as already obtained in \cite{Espriu:2020dge}, where the consequences of the $\mu_5$ corrections for the pole and screening masses, as well as the modification of the pion velocity were analyzed in detail. 

Since the $\mu_5$ modifications to the dispersion relation in \eqref{disprel} are $\Od(p^4)$, as it is the difference $M_\pi^2(T)-M_0^2$,  when setting the external legs on-shell it only affects the LO amplitude. Thus, let us consider the tree level contribution $A_2$ in Fig.\ref{fig:diag}a off the mass-shell, i.e., without specifying the dispersion relation for the external legs, which is given by

\begin{equation}
    A_{2,off-shell} = \frac{3s-\sum_{i=1}^4 p_i^{2}+M_0^{2}}{3F^{2}}
\label{A2off}
\end{equation}
which to leading order, i.e., $p_i^2=M_0^2$ reduces to the well-known Weinberg's low-energy theorem

\begin{equation}
A_2(s)=\frac{s-M_0^2}{F^2}
\label{A2Wein}
\end{equation}

Replacing then the dispersion relation \eqref{disprel} in \eqref{A2off} gives rise to the following $\Od(p^4)$ $\mu_5$-dependent correction to the amplitude:

\begin{equation}
\delta A_{4,dr}(\Sr,\Tr,\Ur;\mu_5)=\frac{1}{3F^{2}}[\tilde{\kappa_{2}}(S_{0}^{2}+T_{0}^{2}+U_{0}^{2})+4(\tilde{\kappa_{1}}-\tilde{\kappa}_{3})M_0^{2}]
\label{A4dr}
\end{equation}

\subsection{LSZ residue}

As explained above, including properly the dressed external lines as given in Fig.\ref{fig:diag}d,e, amounts to consider scattering of pions with the modified dispersion relation \eqref{disprel}. According to the LSZ formalism, we must then amputate those dressed external legs by multiplying the 4-point Green function by the adequate momentum function removing the pole for each leg and extracting the corresponding residue. 

One must therefore repeat the standard LSZ derivation \cite{IZbook} considering asymptotic external fields satisfying a generic dispersion relation  $p_0^2=f(\vert\vec{p}\vert^2)$, which in our present case \eqref{disprel} corresponds at this order to

\begin{equation}
f(\vert\vec{p}\vert^2)=\vert\vec{p}\vert^2 + M_\pi^2(T) -\tilde\kappa_2\vert\vec{p}\vert^2+(\tilde\kappa_3-\tilde\kappa_1-\tilde\kappa_2)M_0^2
\label{fdr}
\end{equation}

Defining as customary the corresponding asymptotic field operator in terms of creation and annihilation operators, now with the modified dispersion relation, i.e. for an asymptotic  scalar field 

\begin{equation}
\varphi(x)=\int \frac{d^3 \vec{k}}{(2\pi)^3 2k_0}  \left[ a(k)e^{-ikx}+a^\dagger (k)e^{ikx} \right],
\end{equation}
satisfying $k_0^2=f(\vert\vec{k}\vert^2)$, one can follow the same steps as in \cite{IZbook} and check that in the end one has just to modify the standard LSZ amputation prescription as

\begin{equation}
\left\{\left\{\prod_{i=1}^n \left[-q^2+M_\pi^2 \right]\right\} G(q_1,\dots,q_n)\right\}_{OS}\rightarrow \left\{\left\{\prod_{i=1}^n \left[-(q^0_i)^2+f(\vert\vec{q}\vert^2)\right]\right\} G(q_1,\dots,q_n)\right\}_{OS}
\label{LSZgen}
\end{equation}
for $n$ external legs, with $G$ the $n$-point connected Green function with all momenta set as incoming and where "OS" means the mass shell condition $(q^0_i)^2=f(\vert\vec{q_i}\vert^2)$. 

Therefore, in our case, from \eqref{LSZgen} and \eqref{fdr}, we are left with the following correction factor with respect to the $\mu_5=0$ case, coming from the amputation of the external legs, at the order we are considering here:

\begin{eqnarray}
\frac{q^{2}-M_\pi^2(T) +\tilde\kappa_2\vert\vec{q}\vert^2-(\tilde\kappa_3-\tilde\kappa_1-\tilde\kappa_2)M_0^2}
{q^{2}-M_\pi^2(T) + \tilde{\kappa_{1}}q^{2} + \tilde{\kappa_{2}}q_0^{2} - \tilde{\kappa_{3}}M_0^2}
&=&\frac{(1-\tilde\kappa_1-\tilde\kappa_2)\left[q^{2}-M_\pi^2(T)\right]+\tilde{\kappa_{1}}q^{2} + \tilde{\kappa_{2}}q_0^{2} - \tilde{\kappa_{3}}M_0^2}
{q^{2}-M_\pi^2(T) + \tilde{\kappa_{1}}q^{2} + \tilde{\kappa_{2}}q_0^{2} - \tilde{\kappa_{3}}M_\pi^2(T)} + \Od(p^6) 
\nonumber\\
&=&1-\tilde\kappa_1-\tilde\kappa_2 + \Od(p^6) 
\end{eqnarray}
being $q$ a particular external four-momentum. As in the dispersion relation modification, the above correction  affects  at this order only the LO amplitude in \eqref{A2Wein}, giving rise to the following $\Od(p^4)$ correction:

\begin{equation}
\delta A_{4,LSZ}(\Sr,\Tr,\Ur;\mu_5)=-4(\tilde\kappa_1+\tilde\kappa_2)\frac{s-M_0^2}{F^2}
\label{A4LSZ}
\end{equation}
where the factor of $4$ comes from the four external legs. 

Therefore, collecting the results in \eqref{A4tree}, \eqref{A4dr} and \eqref{A4LSZ}, the total amplitude is given by:

\begin{equation}
A (\Sr,\Tr,\Ur;T,\mu_5)=A (\Sr,\Tr,\Ur;T,\mu_5=0)+\frac{4\mu_5^2}{3F^{4}} \left\{-3(3\kappa_{1}+4\kappa_{2})s + 3\kappa_{2}S_{0}^{2} + 3\left[4(\kappa_1+\kappa_2)-\kappa_3\right]M_0^2\right\}
\label{totalamp}
\end{equation}

Note that, although for simplicity we have expressed the amplitude in terms of  $M_0,F$, for numerical purposes we will express those parameters in terms of the physical $M_\pi\simeq$ 140 MeV and $F_\pi\simeq$ 93 MeV using the $T=\mu_5=0$ one-loop expressions in  \cite{Gasser:1983yg} relating the physical $M_\pi$ and $F_\pi$ with $M_0,F$. 

\subsection{Partial waves, unitarity and resonance poles}

As we mentioned above, we are going to use the IAM to unitarize the amplitude in order to study poles in the complex plane  associated to resonances. The IAM at $T=\mu_5=0$ has proven a very useful method to unitarize meson scattering partial waves $t^{IJ}$ of well-defined isospin $I$ and total angular momentum $J$ \cite{Dobado:1996ps,GomezNicola:2001as} which in many cases allows to extend the ChPT applicability range and in addition it generates resonances in the proper channels as poles in the appropriate Riemann sheets in the $s$-complex plane. In particular, for $\pi\pi$ scattering, it describes the poles corresponding to the $f_0(500)/\sigma$ in the $I=J=0$ channel and the $\rho (770)$ for $I=J=1$.

The extension of this method to nonzero $T$ has allowed to describe successfully the thermal poles of the $f_0(500)$ and $\rho(770)$ for $\pi\pi$ scattering \cite{Dobado:2002xf} as well as the $K_0^* (700)$ and $K^* (890)$ for $\pi K$ scattering \cite{GomezNicola:2020qxo,GomezNicola:2023rqi}. A very interesting feature in that context is that the thermal pole of the $f_0(500)$, within a saturation approach, describes the scalar susceptibility and hence the transition temperature $T_c$  consistently with lattice results \cite{Ferreres-Sole:2018djq}. We will be more specific about that in section \ref{sec:Tcmu5}. Similar  conclusions hold for the $K_0^* (700)$ and the $I=1/2$ scalar susceptibility, which involves also $U(1)_A$ restoration 
\cite{GomezNicola:2020qxo,GomezNicola:2020qxo}. 

At $\mu_5\neq 0$ and $T\neq 0$, we define the partial waves following the standard conventions  \cite{GomezNicola:2001as,GomezNicola:2002tn} i.e., the isospin projection reads

\begin{eqnarray}
    T^0(\Sr,\Tr,\Ur;T,\mu_5) &=& 3A(\Sr,\Tr,\Ur;T,\mu_5)+A(\Tr,\Sr,\Ur;T,\mu_5)+A(\Ur,\Tr,\Sr;T,\mu_5) \nonumber\\
    T^1(\Sr,\Tr,\Ur;T,\mu_5) &=& A(\Tr,\Sr,\Ur;T,\mu_5)-A(\Ur,\Tr,\Sr;T,\mu_5) \nonumber \\
    T^2(\Sr,\Tr,\Ur;T,\mu_5) &=& A(\Tr,\Sr,\Ur;T,\mu_5)+A(\Ur,\Tr,\Sr;T,\mu_5)
\label{isospin}
\end{eqnarray}
after which one can set the center of mass conditions to construct the projection into $IJ$ partial waves for $\pi\pi$ scattering:

\begin{equation}
    t^{IJ}(s;T,\mu_5)=\frac{1}{64\pi}\int_{-1}^{1}dx P_{J}(x)T^{I}\left[S_{0}=\sqrt{s},\vec{S}=\vec{0},T_{0}=0,|\vec{T}|^{2}=-t(s,x),U_{0}=0,|\vec{U}|^{2}=-u(s,x)\right]
\label{partialwaves}
\end{equation}

where $P_J$ is the $J$-th Legendre polynomial and 

\begin{align}
 t(s,x)=-\frac{s\sigma^2(s)}{2}(1-x), \quad u(s,x)=-\frac{s\sigma^2(s)}{2}(1+x) 
\label{CM}
\end{align}
with $\sigma(s)=\sqrt{1-\frac{4M_{\pi}^{2}}{s}}$  the phase space of two identical particles of mass $M_{\pi}$.

From \eqref{totalamp}, the 
$\mu_{5}$ corrections to the lowest angular momentum perturbative ChPT partial waves read:

\begin{eqnarray}
    \Delta t ^{00}&=&-\frac{\mu_5^2}{8\pi F^{4}}[\kappa_{1}'s+\kappa_{2}'M_0^2],
     \label{t00corr}\\
    \Delta t ^{11}&=&-\frac{\mu_5^2}{24\pi F^{4}}(s-4M_0^{2})\kappa_{3}',\label{t11corr}\\
    \Delta t^{20}&=&\frac{\mu_5^2}{8\pi F^{4}}[\kappa_{4}'s-2M_0^{2}\kappa_{5}'],
\label{t20corr}
\end{eqnarray}
with the following combinations of $\kappa_i$ constants 

\begin{align}
\kappa_{1}'&=6\kappa_{1}+5\kappa_{2},&\kappa_{2}'&=-8\kappa_{1}-4\kappa_{2}+5\kappa_{3},&
\kappa_{3}'&=3\kappa_{1}+4\kappa_{2},\nonumber\\
\kappa_{4}'&=3\kappa_{1}+4\kappa_{2},&
\kappa_{5}'&=2\kappa_{1}+4\kappa_{2}+\kappa_{3}.
 \label{kappaprime}
 \end{align}

Note that the $I=J=1$ channel depends only on one $\kappa_i$ combination, unlike the scalar partial waves,  since it vanishes at the two-pion threshold due to its vector nature and therefore the $s$ and $-4M_0^2$ coefficients coincide. 

The ChPT partial waves satisfy the unitarity conditions only perturbatively, i.e., if we write $t=t_2+t_4+\dots$ for a given $IJ$, we have 

\begin{equation}
\im t_2(s;T,\mu_5)=0, \quad \im t_4(s;T,\mu_5)=\sigma_T (s;T) t_2^2(s;T,\mu_5) \quad (s\geq 4M_\pi^2)
\label{pertunit}
\end{equation}
where $\sigma_T(s;T)=\sigma(s)\left[1+2n_B(\sqrt{s};T)\right]$ with $n_B(x;T)=\left[e^{x/T}-1\right]^{-1}$ the Bose-Einstein distribution function, is the so called thermal phase space, which accounts for scattering processes inside the thermal bath \cite{GomezNicola:2002tn,GomezNicola:2023rqi}. Recall that $t_2$ is independent of $T$ and $\mu_5$ and the $\mu_5$ contributions to $t_4$ are real for real $s$. Therefore, they do not affect the  thermal perturbative unitarity relation \eqref{pertunit}, which is the  perturbative version of the unitarity relation for partial waves

\begin{equation}
\im t^{IJ} (s;T,\mu_5)=\sigma_T(s;T) \vert t^{IJ}(s;T,\mu_5) \vert^2 \quad (s\geq 4M_\pi^2)
\label{unit}
\end{equation}
which can also be written in terms of the inverse amplitude as $\im (1/t^{IJ})=-\sigma_T$.

Unitarization methods aim to construct amplitudes satisfying exactly unitarity, which in particular allow to generate dynamically physical resonances. The difference between the various methods lies in the approximation performed for $\re (1/t)$. We will follow here the IAM, for which $\re(1/t)$ is approximated by the ChPT amplitude to fourth order. Including the $\mu_5$ dependence, the unitarized IAM amplitude  is given for each partial wave by

\begin{equation}
t_U (s;T,\mu_5) = \frac{\left[ t_2 (s) \right]^2}{t_2(s)-t_4(s;T,\mu_5)}
\label{iam}
\end{equation}

We will use the above IAM formula to generate the $f_0(500)$ and $\rho (770)$ resonances in the $I=J=0$ and $I=J=1$ respectively, which appear as poles in the second Riemann sheet of the amplitude in the $s$ complex plane across the unitarity cut $s\geq 4M_\pi^2$. As customary, we parametrize the pole position as $s_p=(M_p-i\Gamma_p/2)^2$ so that $M_p,\Gamma_p$ would correspond to the mass and width for the case of narrow resonances $\Gamma_p\ll M_p$, as in the $\rho$ case.  

As mentioned above, at the order we are considering here,  $\mu_5$ appears as a correction to $t_4$ not affecting the perturbative unitarity relation \eqref{pertunit}. Higher order $\mu_5$ corrections could yield nontrivial corrections through the modification of vertices and/or the dispersion relation, but since the distribution functions of pions  in the thermal bath is not affected by $\mu_5$, and neither are then the relevant scattering processes inside the thermal bath 
\cite{GomezNicola:2023rqi}, 
the unitarity relation \eqref{unit} should keep the same structure.

\section{The scalar susceptibility and the $\mu_5$ dependence of the transition temperature}
\label{sec:Tcmu5}

A key observable signaling chiral symmetry restoration is the light scalar susceptibility

\begin{eqnarray}
    \chi_{S}(T,\mu_5)=-\frac{\partial}{\partial m}\langle\bar{q}q\rangle_{l}(T,\mu_5)=
    \int_{T}dx[\langle\bar{q}_{l}q_{l}(x)\bar{q}_{l}q_{l}(0)>-\langle\bar{q}q\rangle_{l}^{2}(T,\mu_5)]
\label{chisdef}
\end{eqnarray}
with $\condl=\langle \bar u u+ \bar d d\rangle$ the light quark condensate. The susceptibility $\chi_S$  develops a peak at the transition temperature in the crossover regime that should get stronger as the light $u,d$ quark masses decrease, becoming divergent in the light chiral limit if the transition is of second order \cite{Smilga:1995qf,Aoki:2009sc,Bazavov:2011nk,Bazavov:2018mes,Ding:2019prx,Ratti:2018ksb,Bazavov:2019lgz,Guenther:2020jwe,Nicola:2020smo}. 

That quantity has, by definition, the quantum numbers of the vacuum, i.e., those of the $I=J=0$ $f_0(500)$ state. In fact,  it has been shown that it is a very good approximation to saturate the light scalar susceptibility with the $f_0(500)$ thermal resonance near the transition \cite{Nicola:2013vma,Ferreres-Sole:2018djq} which yields a peak for $\chi_S(T)$ fully compatible with lattice results. Such saturation approach reads in our present case

\begin{equation}
\frac{\chi_S(T,\mu_5)}{\chi_S(0,\mu_5)}=\frac{M_S^2(0,\mu_5)}{M_S^2(T,\mu_5)}
\label{chis}
\end{equation}
where we have included the $\mu_5$ dependence and  $M_S^2(T,\mu_5)=\mbox{Re} \left[s_p(T,\mu_5)\right]=M_p^2(T,\mu_5)-\Gamma_p^2(T,\mu_5)/4$ behaves as the thermal mass of the $f_0(500)$. Note that $\chi_S$ scales as the inverse of the squared mass of the lightest scalar state, since it is nothing but the $\bar q q$ correlator at vanishing four-momentum, according to \eqref{chisdef}.

The previous approach will allow us to obtain the dependence $T_c(\mu_5)$ as the position of the $\chi_S$ peak, and compare it with lattice results and with the previous work \cite{Espriu:2020dge} where it was determined from the vanishing point of the light quark condensate $\condl (T,\mu_5)$, calculated from the ChPT free energy. We emphasize that determining $T_c$ from the thermal $f_0(500)$ saturated unitarized approach is closer 
 to the lattice results and to the  QCD transition since the dominant degree of freedom for that observable near the transition is indeed the thermal $f_0(500)$ \cite{Ferreres-Sole:2018djq}. Actually, as we recall in detail in section \ref{sec:num}, at $\mu_5=0$ this approach not only provides the expected peak behaviour consistently with lattice data, but the value of $T_c$ obtained using  the same LEC describing $T=0$ physics is remarkably close to the lattice determinations. Conversely, the ChPT quark condensate only describes the qualitative chiral restoring decreasing behaviour, but the value of $T_c$ obtained merely with the pion component lies well above the lattice one and  actually  one needs to include many heavier hadrons to reduce that value to the expected one  \cite{Jankowski:2012ms}. Nevertheless, it was shown in  \cite{Espriu:2020dge} that the ratio $T_c(\mu_5)/T_c(0)$ of lattice results can be very reasonably described just with the ChPT expressions for the quark condensate, yielding a decent determination at least of the $\kappa_i$ combination surviving in the chiral limit. More details will be provided in section \ref{sec:fit}.

\section{Numerical Results}
\label{sec:num}

\subsection{$T_c(\mu_5)$ and fits of $\kappa_i$  to lattice}
\label{sec:fit}

As mentioned above, the undetermined new LEC $\kappa_i$ appearing in the $\Od(p^4)$ $\mu_5\neq 0$ lagrangian \eqref{L4mu5} were fitted in \cite{Espriu:2020dge} to lattice measurements. In particular, to the available lattice data for $T_c(\mu_5)$ in \cite{Braguta:2015zta} and to the chiral charge density $\rho_5$ and topological susceptibility $\chi_{top}$ in  \cite{Astrakhantsev:2019wnp}. 

Regarding $T_c$, the fit in \cite{Espriu:2020dge} was performed with the NNLO ChPT quark condensate $\condl$, which depends on the combinations $\kappa_a=2\kappa_1-\kappa_2$ and $\kappa_b=\kappa_1+\kappa_2-\kappa_3$, where $\kappa_b$ is the constant multiplying $M_\pi^2$ in $\condl$.  Even though the lattice results correspond to a high value of the pion mass,  the fit in \cite{Espriu:2020dge} for the ratio $T_c(\mu_5)/T_c(0)$   with the quark condensate up to $\mu_5=475$ MeV yields  a quite reasonable $\chi^2/dof\simeq 1.4$, reproducing consistently the increasing $T_c(\mu_5)$ behaviour. Actually, such fits are almost insensitive to $\kappa_b$ and in fact one gets a very good fit with the chiral limit expressions, showing little dependence on the pion mass for that ratio and providing a good determination for $\kappa_a=(2.3\pm 0.4) \times 10^{-3}$, with a reasonably small uncertainty given the small number of lattice points available.  

On the other hand, $\kappa_3$ can be fitted independently from the  ratio $\chi_{top}(\mu_5)/\chi_{top}(0)$, which at this order in ChPT depends solely on that constant.   Although in that case  $\chi^2/dof\simeq 1.1$ for the best fit is still quite good, the obtained $\kappa_3=(0.5\pm 0.9) \times 10^{-3}$ shows a larger uncertainty. Since $\kappa_{1,2,3}$ do not enter any other lattice observable at that order, their individual determination involved large uncertainties with the analysis performed in \cite{Espriu:2020dge}. In Table \ref{tab:kappa} we provide the estimate given in \cite{Espriu:2020dge} for the individual $\kappa_{1,2,3}$ values combining all the  information obtained from the fits in that paper. 

 Our present analysis provides new interesting possibilities for determining the $\kappa_{1,2,3}$ constants from lattice results. Thus, on the one hand, as commented, the prediction for $T_c(\mu_5)$ from the peak of the saturated scalar susceptibility is much more reliable as a truly transition temperature than the prediction from the ChPT quark condensate, as confirmed by the numerical values for $T_c$ given below. Besides, note that in the lattice works \cite{Braguta:2015zta,Braguta:2015owi}  $T_c(\mu_5)$ is determined precisely from the peak of the scalar susceptibility. On the other hand, $\Delta t^{00}$ in \eqref{t00corr} and hence $T_c$ depend on  the combinations $\kappa'_1$ and $\kappa'_2$ in \eqref{kappaprime}, which are different from $\kappa_{a,b}$ appearing in the quark condensate. Thus,  a fit of those combinations combined with the analysis  in \cite{Espriu:2020dge} should allow   for a better individual determination of  $\kappa_{1,2,3}$.

The pion scattering amplitude and poles depend in addition on the LEC $\bar l_{1,2,3,4}$ of the $\mu_5=0$ lagrangian in \cite{Gasser:1983yg}. We will take here the values of $\bar l_{1,2}$ LEC fitted in \cite{Hanhart:2008mx} to scattering data, namely $\bar l_1=-0.1\pm 0.2$, $\bar l_2=5.8\pm 0.2$. For $\bar l_3$ we take the estimate $\bar l_3=3.4\pm 0.8$  in \cite{FlavourLatticeAveragingGroup:2019iem} for the sake of an easier comparison with the results in  \cite{Espriu:2020dge}, where $\bar l_3$ was fixed to that value and it was the only one of the $\bar l_i$ showing up in the fits. That $\bar l_3$ value is fully compatible with the value used in  \cite{Hanhart:2008mx}, which is not fitted but taken directly from \cite{Gasser:1983yg}. For $\bar l_4$ we take $\bar l_4=4.4\pm 0.9$, the same value as in  
 \cite{Hanhart:2008mx} and \cite{Gasser:1983yg} whose central value coincides with the estimate in  \cite{FlavourLatticeAveragingGroup:2019iem}. 
 
 With the above $\bar l_i$ set, we obtain for the peak position of the saturated scalar susceptibility \eqref{chis}, $T_c(0)=158.5$ MeV, which, as advanced, is remarkably close to the lattice prediction $T_c=156\pm 1.5$ MeV \cite{Guenther:2020jwe}, the uncertaintites of the LEC yielding compatibility with lattice results \cite{Ferreres-Sole:2018djq}. 

As a first analysis, we have fitted only the critical temperature obtained using the saturated scalar susceptibility leaving $\kappa_1'$ and $\kappa_2'$ as fit parameters. We have considered the lattice points provided in \cite{Braguta:2015zta} for the critical temperature ratio. In the physical limit, the sensitivity of the fit to $\kappa_1'$ turns out to be much stronger than to the constant proportional to the pion mass, i.e. $\kappa_2'$. Indeed, trying to fit the lattice data, one can see that the error in $\kappa_2'$ is very large. In that sense, the behaviour of our present theoretical result is similar to that in \cite{Espriu:2020dge}  based on the quark condensate and one could think  of fitting the lattice points with the chiral limit result only, as in that work. 

However, there is an important qualitative difference between the two approaches regarding the chiral limit. The ChPT quark condensate merely reduces its value in that limit, and consequently $T_c$ is also reduced, but the qualitative form of the curve is similar and, as mentioned, the ratio   $T_c (\mu_5)/T_c(0)$ is almost unaffected. In fact, since it is based on the ChPT perturbative approach, it does not reproduce the expected inflection point at $T_c$ in the physical limit. However, the susceptibility approach we are discussing here, based on thermal scattering, resembles more accurately the expected shape, i.e., a peak at $T_c$ in the physical limit, turning into a divergence in the chiral limit at $T_c^{chiral}(\mu_5=0)=125.5$ MeV, also quite close to the lattice predictions \cite{Ding:2019prx}. Such strong qualitative change in the chiral limit is expected to show somehow in the  sensitivity of the fit. This in indeed the case: if we perform the fit in the chiral limit we obtain $\kappa_1'=(7.7\pm 0.8)\times 10^{-3}$ and a $\chi^2/\text{dof}=3.90$, where the uncertainty corresponds to the $95\%$ confidence level of the fit. This result is compatible with  $\kappa_1'=(2.3\pm 7.8)\times 10^{-3}$ estimated from $\kappa_{1,2,3}$ in \cite{Espriu:2020dge} because  of the large error in $\kappa_i$. However, the mean values are quite far from each other. Thus, we conclude that although the dependence with $\kappa_{1}'$ is larger than with $\kappa_{2}'$, it is crucial to consider the mass of the pion, and thus the constant $\kappa_2'$, to accurately determine the critical temperature and fit the selected lattice points in the  approach based on the saturated scalar susceptibility. As we show below, doing so we will obtain a better fit and a  better agreement between the present results and those in \cite{Espriu:2020dge}.

Therefore, in order to reduce the error of the $\kappa_2'$ constant and to  incorporate further lattice results, we will perform a combined fit with $\kappa_{1,2,3}$ as fit parameters and where the fitted observables will be, on the one hand,   $T_c (\mu_5)/T_c(0)$ with the two methods already discussed, i.e., the ChPT condensate from   \cite{Espriu:2020dge} and the scalar susceptibility obtained here, and, on the other hand, the topological susceptibility $\chi_{top}$, with the lattice results in \cite{Astrakhantsev:2019wnp} fitted with the theoretical ChPT analysis in \cite{Espriu:2020dge}. We have used the full $M_\pi\neq 0$ ChPT expression in \cite{Espriu:2020dge} for the condensate and we have checked that using the chiral limit expression instead does not alter significatively the obtained $\kappa_i$ parameters nor their uncertaintites, as expected from our previous comments.

\begin{figure}[h!]
\centering
\vspace{-0.5cm}
\includegraphics[width=0.6\textwidth]{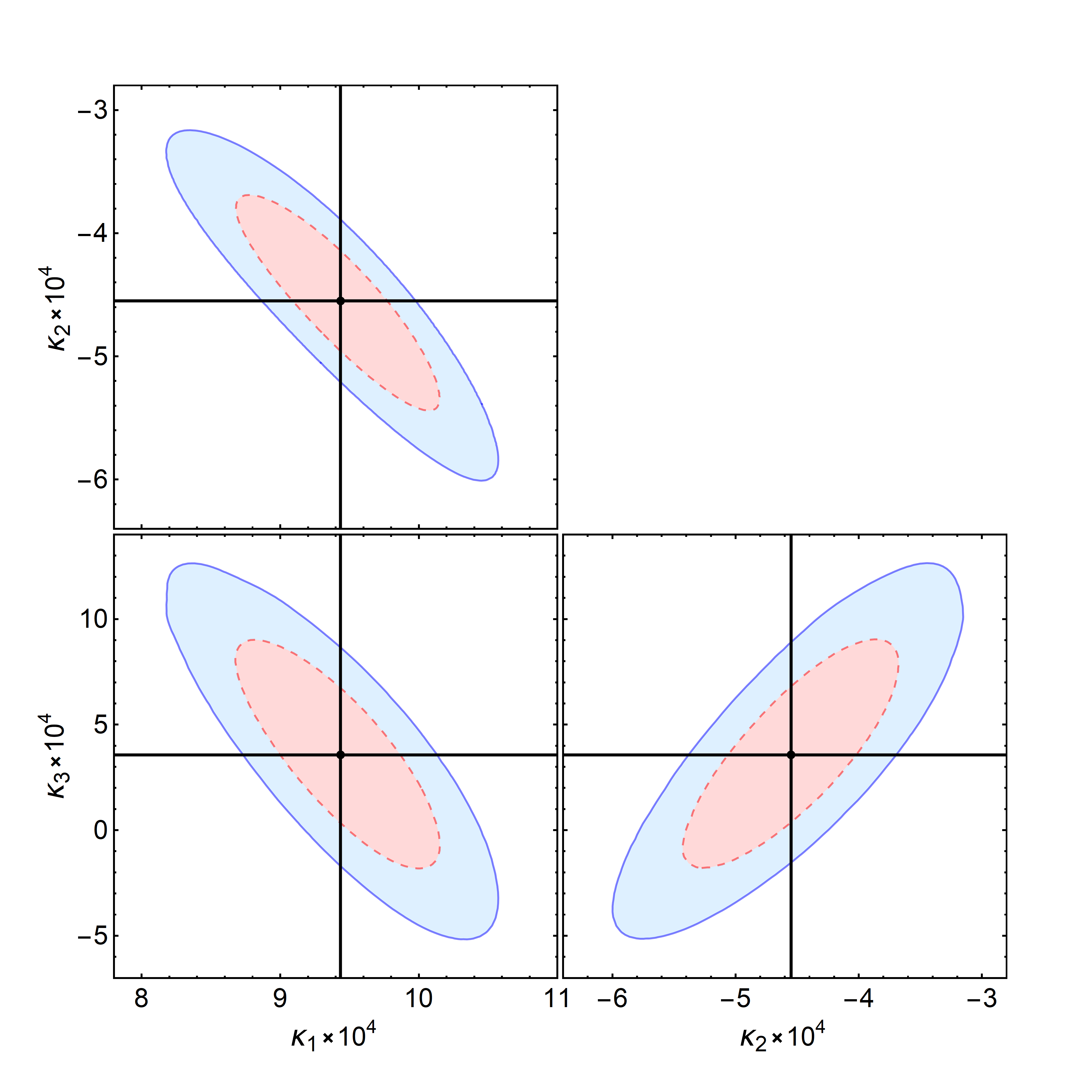}
\vspace{-0.5cm}
	\caption{Error contours for the fit discussed in the main text. The black lines show the ‘best fit’ locations and the contours correspond to the $68\%$ (red dashed line) and $95\%$ (blue full line) confidence levels.}
	\label{fig:Tcfit}
\end{figure}

\begin{figure}[h!]
    \centering
\includegraphics[width=0.48\textwidth]{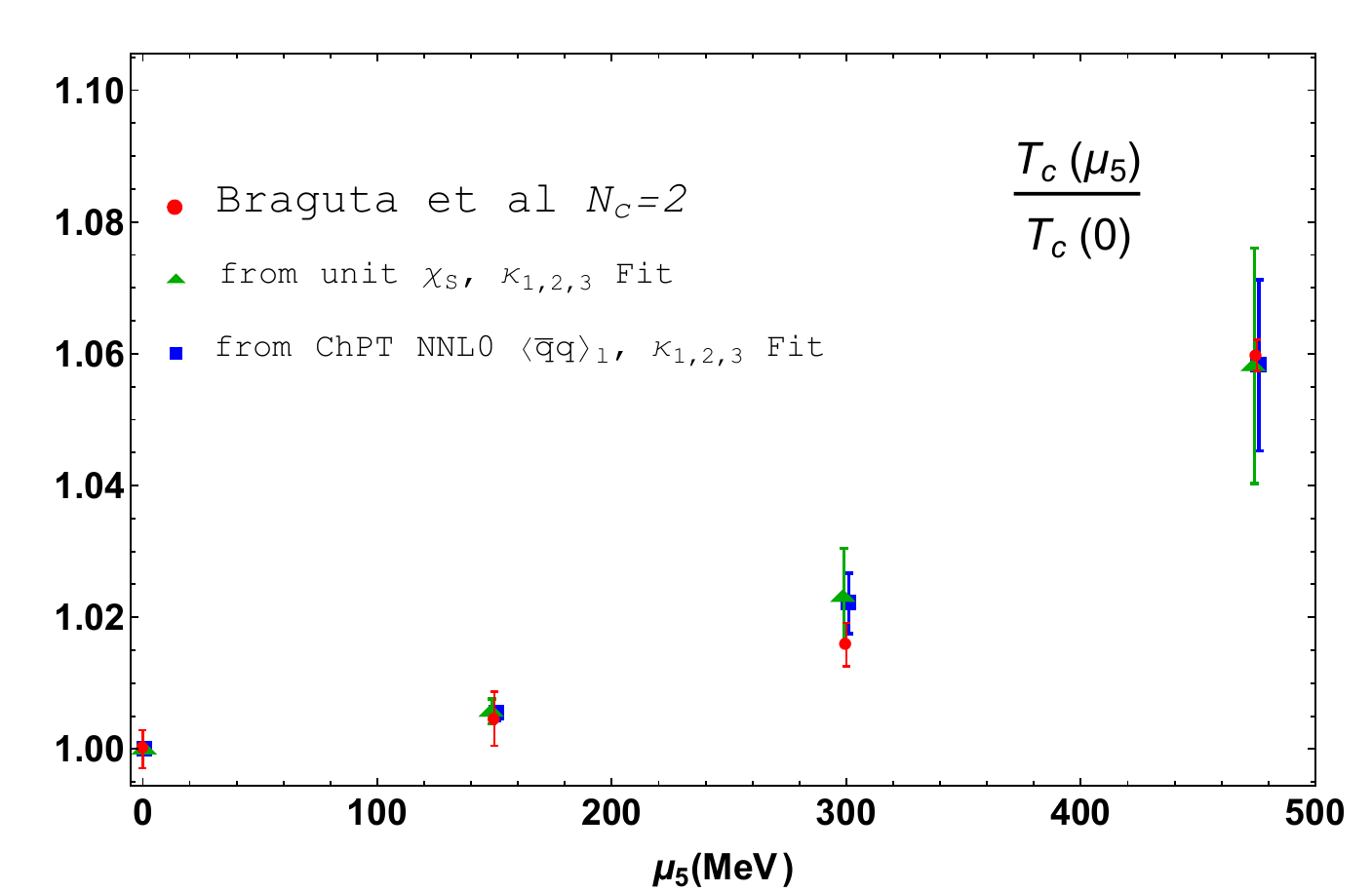}
\includegraphics[width=0.48\textwidth]{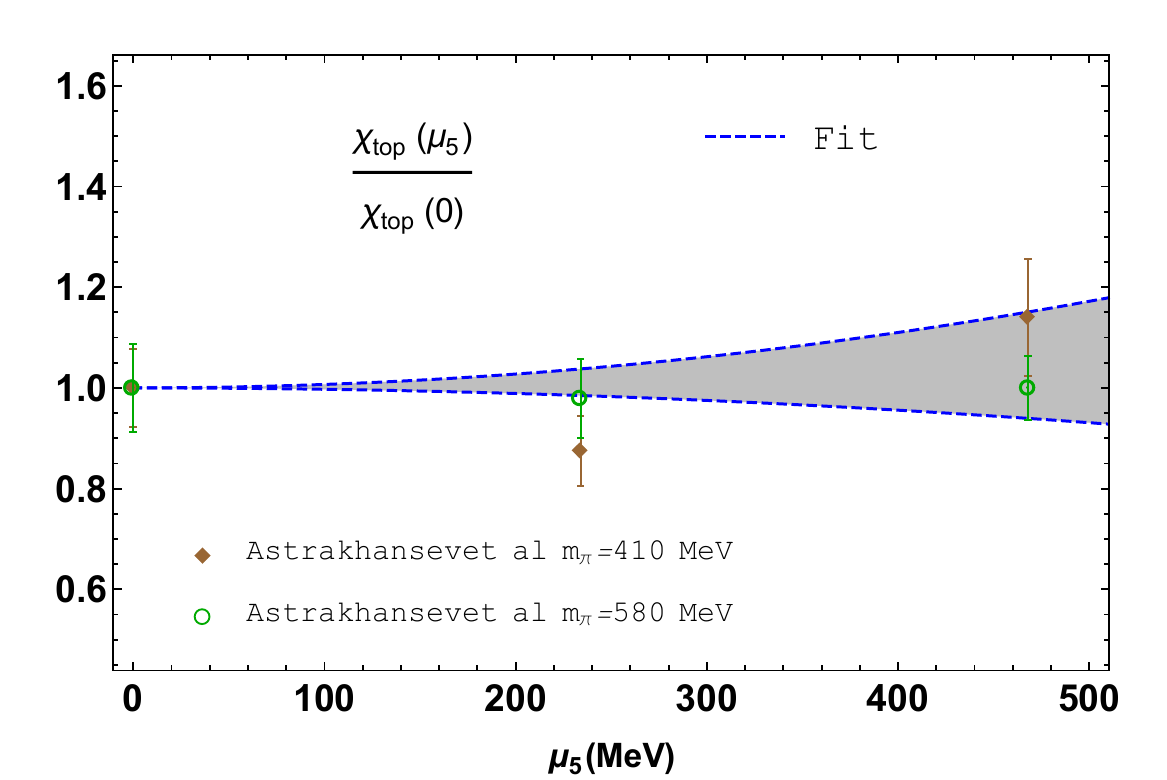}
    \caption{Combined fit of  $T_c (\mu_5)/T_c(0)$ (left) determined from the NNLO ChPT quark condensate \cite{Espriu:2020dge} and the present analysis using the saturated scalar susceptibility, and the topological susceptibility (right) with the ChPT result \cite{Espriu:2020dge}. We also show the lattice results from \cite{Braguta:2015zta}, \cite{Astrakhantsev:2019wnp} and their uncertainties. All the results are plotted with the $\kappa_i$ set corresponding to our present combined fit, given in Table \ref{tab:kappa}.
   }
  \label{fig:FitTc}
\end{figure}

The result of such combined fit is showed in Table \ref{tab:kappa}, where we provide the $\kappa_i$ and their uncertainties for a $95\%$ confidence level.  
The error contours obtained are shown in Fig.\ref{fig:Tcfit} and in Fig.\ref{fig:FitTc} we display  the results of the fit for the $T_c$ ratio and $\chi_{top}$. Lattice points are perfectly compatible with both $T_c$ determinations and confirm the growing behaviour of $T_c(\mu_5)$. In fact, we have checked that $T_c(\mu_5)$ also grows within the same range when using the present $T_c$ determination but with the central $\kappa_i$ values in  \cite{Espriu:2020dge}.  As mentioned, the $T_c$ determination is quite different in both approaches. Namely, $T_c^{ChPT \bar q q}(0)=301$ MeV and $T_c^{latt}(0)=195.8$ MeV for the lattice work   we are considering \cite{Braguta:2015zta}, which, due to the higher pion mass, is notably  higher than the standard value measured in the physical limit \cite{Guenther:2020jwe}.  The error band of the topological susceptibility is similar to that reported in \cite{Espriu:2020dge}, since the uncertainty of $\kappa_3$ obtained from our present fit is indeed of the same order as their value, as it can be seen in table \ref{tab:kappa}.

\begin{table}[h]
\centering
\setlength{\extrarowheight}{6pt}
\begin{tabular}{|p{3.5cm}|c|c|c|p{1.5cm}|}
\hline
  & $\kappa_1\times 10^4$ & $\kappa_2\times 10^4$ & $\kappa_3\times 10^4$ & $\chi^2/\text{dof}$\\[0.2cm]
\hline
 This work (combined fit) & $9.4^{+1.1}_{-1.3}$ & $-4.5^{+1.5}_{-1.4}$ & $3.6^{+9.1}_{-8.7}$ & 1.37\\[0.2cm]
 \hline
Estimate in \cite{Espriu:2020dge} from $T_c$ and $\chi_{top}$ fits& $8\pm 10$ & $-5\pm 10$ & $3\pm 10$ & \parbox{1.5cm}{1.4 ($T_c$)  1.1 ($\chi_{top}$)} \\[0.2cm]
\hline
\end{tabular}
    \caption{Numerical values of the $\kappa_i$ parameters with the fit discussed in the main text. We also display for comparison the estimated values obtained in \cite{Espriu:2020dge}.}
    \label{tab:kappa}
\end{table}

Finally, let us comment that, as discussed in \cite{Espriu:2020dge}, certain physical conditions impose  some sign restrictions on the $\kappa_i$ parameters. 
\color{black} 
Thus, from the NLO ChPT dispersion relation \eqref{disprel} one infers, on the one hand, that to ensure that the pion velocity remains below the speed of light for any value of $\mu_5$, $\kappa_2$ must be negative. On the other hand,  the combination $\kappa_1-\kappa_3$ must be positive to make sure that the squared pion mass at NLO remains positive for all values of $\mu_5$, i.e., that pions do not become tachyonic. The fit value we have obtained here for $\kappa_2$, provided in Table \ref{tab:kappa} remains negative within uncertainties, while the central value of $\kappa_1-\kappa_3\sim (6\pm 9)\times 10^{-4}$ is positive, although the larger $\kappa_3$ uncertainty affects the sign of that combination. Additional lattice data for the topological susceptibility would be needed to achieve a smaller uncertainty, since this is the  observable most sensitive to $\kappa_3$, as explained.

\subsection{Phase shifts, resonances and the scalar susceptibility}

Here, we present our results for different observables, using the values of $\kappa_i$ obtained in the fit performed in section \ref{sec:fit} and provided in Table \ref{tab:kappa}.  The uncertainty bands in the different observables coming from those of $\kappa_i$ in Table \ref{tab:kappa}, grow with $\mu_5^2$. Thus, the upper value considered here $\mu_5=475$ MeV sets a natural applicability limit of our approach, since for that value the uncertainty bands start to overlap with the $\mu_5=0$ curves.

\begin{figure}[h!]
    \centering
    \includegraphics[width=0.45\textwidth]{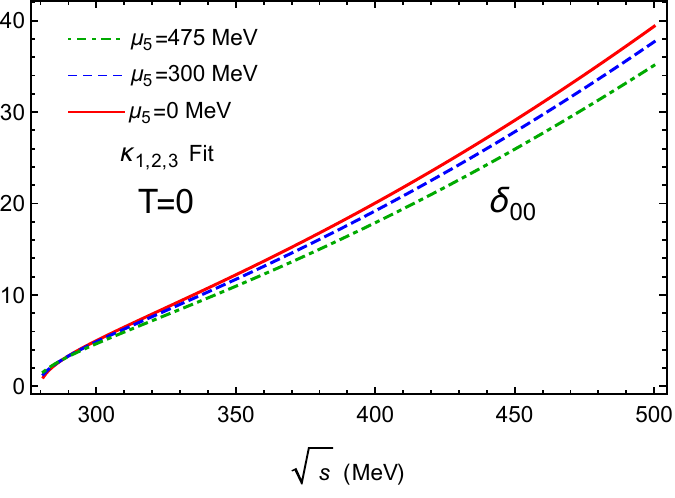}\hspace{0.5cm}
     \includegraphics[width=0.45\textwidth]{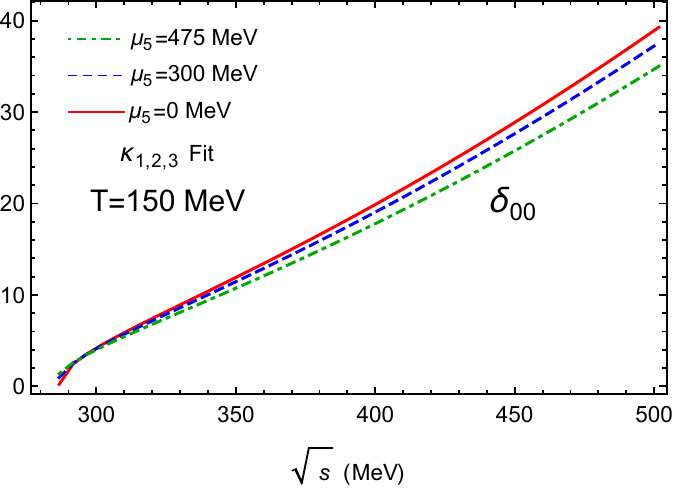}\hspace{0.5cm}
    \includegraphics[width=0.45\textwidth]{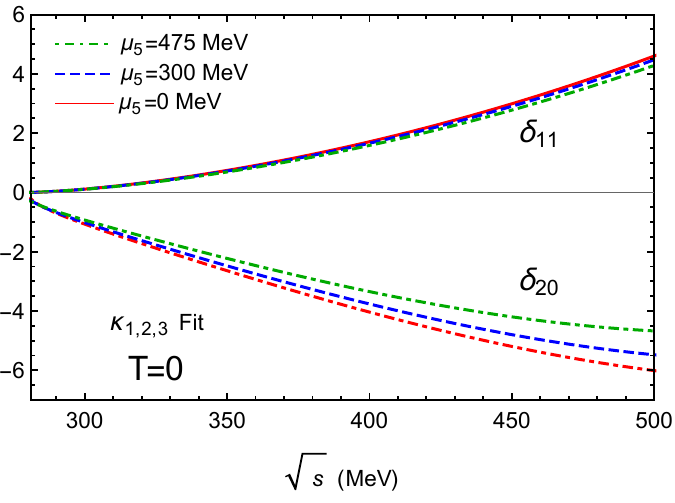}
     \includegraphics[width=0.45\textwidth]{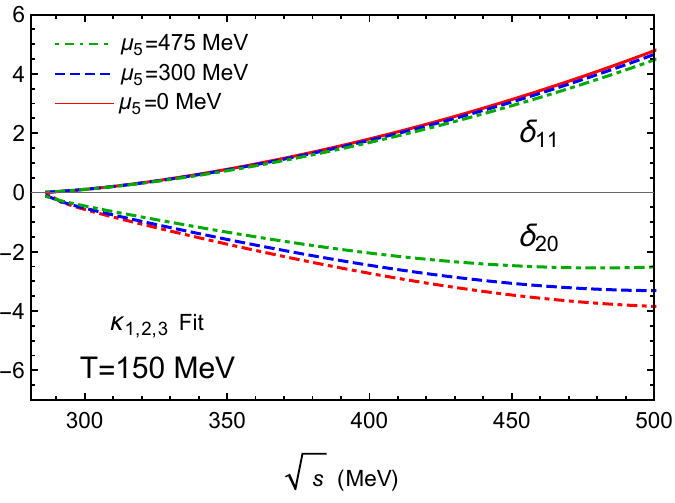}
    \caption{$\mu_5$ dependence of the perturbative phase shifts $\delta_{IJ}$ for $IJ=00,\,11,\,20$ at $T=0$ and $T=150$ MeV.}
  \label{fig:desmu5pert}
\end{figure}

\begin{figure}[h]
    \centering
    \includegraphics[width=0.45\textwidth]{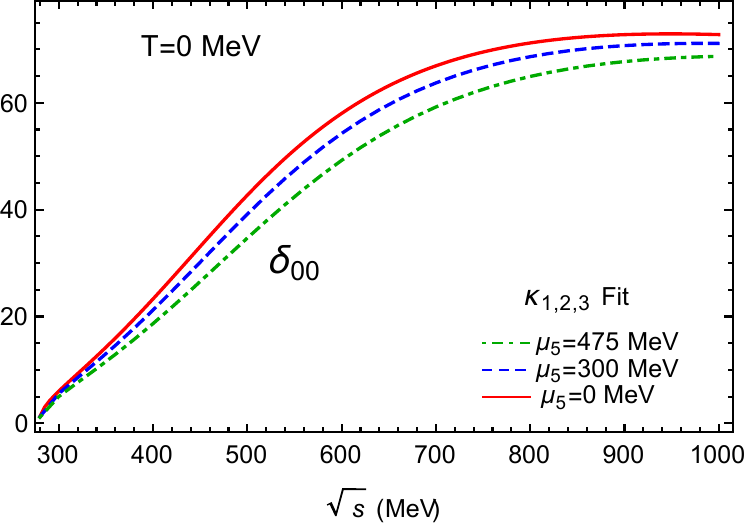}\hspace{0.5cm}
    \includegraphics[width=0.45\textwidth]{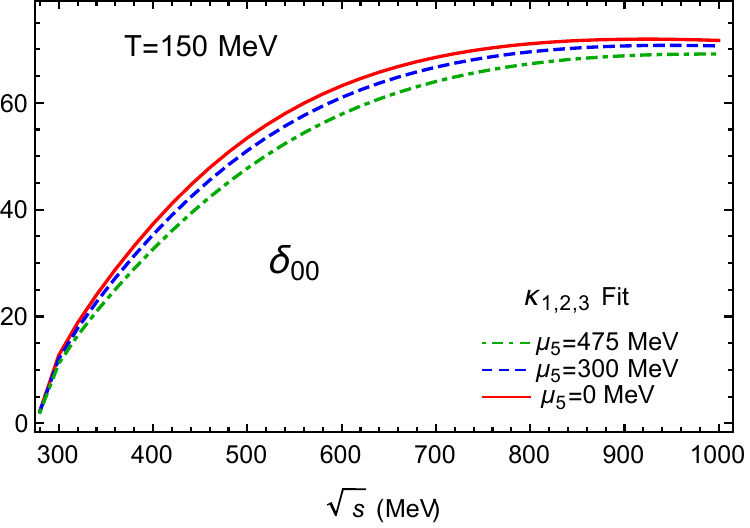}
    \includegraphics[width=0.45\textwidth]{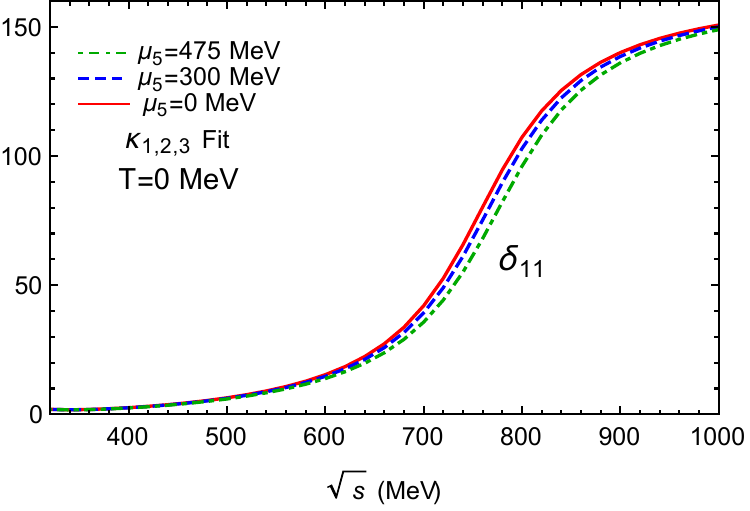}\hspace{0.5cm}
    \includegraphics[width=0.45\textwidth]{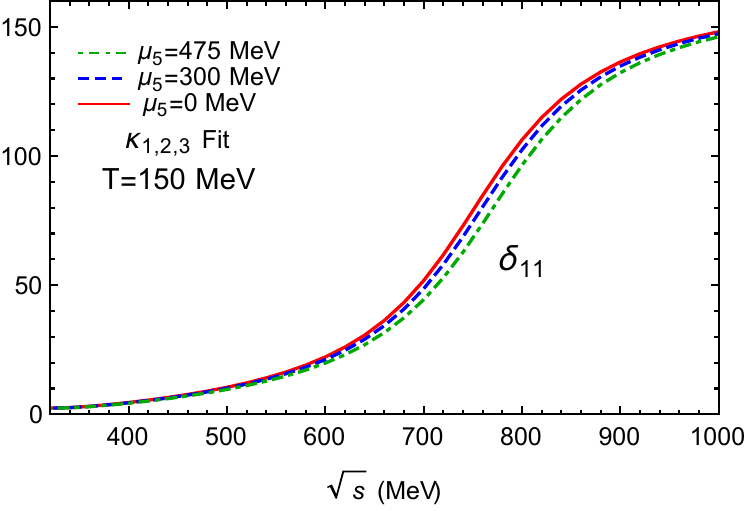}
    \caption{$\mu_5$ dependence of the unitarized phase shifts $\delta_{IJ}$ for $IJ=00,\,11$ at $T=0$ and $150$ MeV.}
  \label{fig:desmu5unit}
\end{figure}

The scattering phase shifts $\delta_{IJ}$ in different channels for $\mu_5=0,300$, $475$ MeV, are displayed in  Fig.\ref{fig:desmu5pert} for the perturbative amplitude and in Fig.\ref{fig:desmu5unit} for the unitarized one in the  resonant channels, where we appreciate the typical Breit-Wigner shape in the $I=J=1$ around the $\rho$ mass. In both cases we represent the results for $T=0$ and $T=150$ MeV.  It is noteworthy that each channel retains its respective attractive or repulsive nature with $\mu_5\neq 0$ and the absolute value of all the phase shifts is reduced as $\mu_5$ increases. Such reduction goes in the opposite direction as the temperature effects \cite{GomezNicola:2002tn}. We  also see that the $\mu_5$ variation in the vector-isovector channel is quite small.

\begin{figure}[h!]
    \centering
    \includegraphics[width=0.45\textwidth]{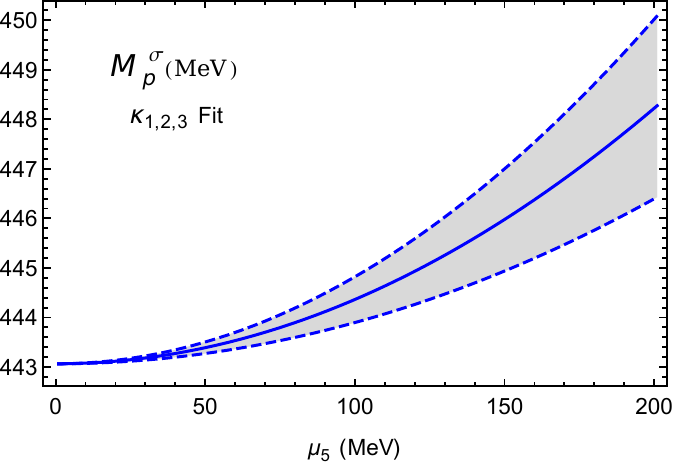}\hspace{0.5cm}
    \includegraphics[width=0.45\textwidth]{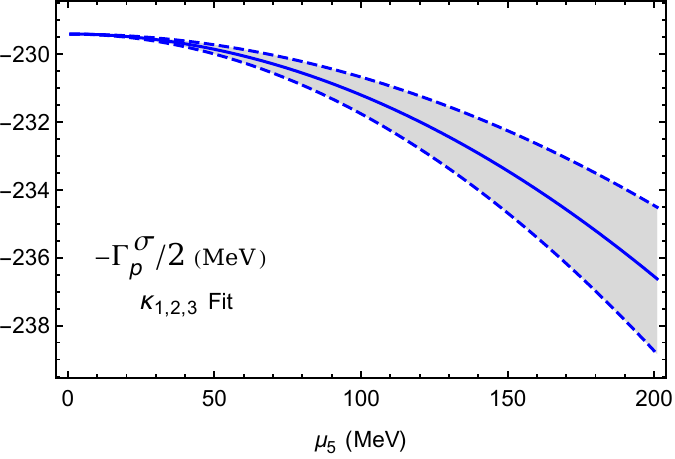}\\
        \includegraphics[width=0.45\textwidth]{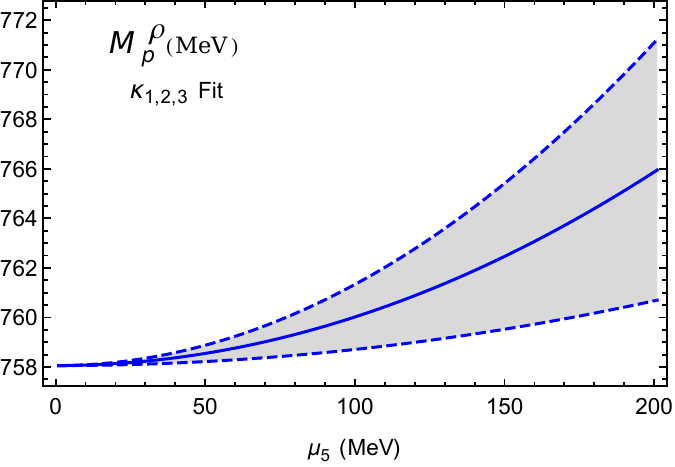}\hspace{0.5cm}
    \includegraphics[width=0.45\textwidth]{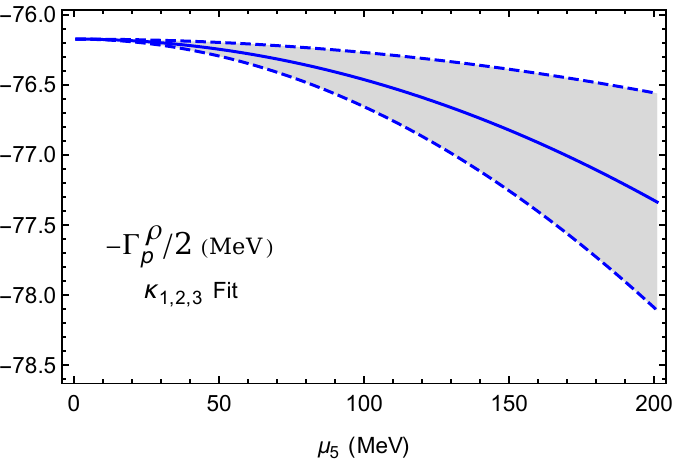}
    \caption{Corrections to the $M_{p}$ and 
 $\Gamma_{p}$ parameters of the $f_{0}(500)$ and $\rho(770)$ resonances due to $\mu_{5}$ at $T=0$.}
\label{fig:Massandwidthsigmarhomu5evT0}
\end{figure}

Our results for the pole parameters as a function of $\mu_5$ for $T=0$ are displayed in Fig.\ref{fig:Massandwidthsigmarhomu5evT0}. The  uncertainty bands corresponding to those of the $\kappa_i$  in Table \ref{tab:kappa} are showed, confirming, as mentioned above, that our results are predictive for low and moderate values of $\mu_5$. It is worth pointing out that in general, we expect a much softer dependence with the fourth-order LEC for the pole in the $I=J=0$ channel than in the $I=J=1$ one \cite{Pelaez:2015qba}. That explains the narrower uncertainty bands for the $f_0(500)/\sigma$ in Fig.\ref{fig:Massandwidthsigmarhomu5evT0} even though the $I=J=1$ amplitude depends only on one single $\kappa_i$ combination, as given in \eqref{t11corr}.

\begin{figure}[h!]
    \centering
\includegraphics[width=0.45\textwidth]{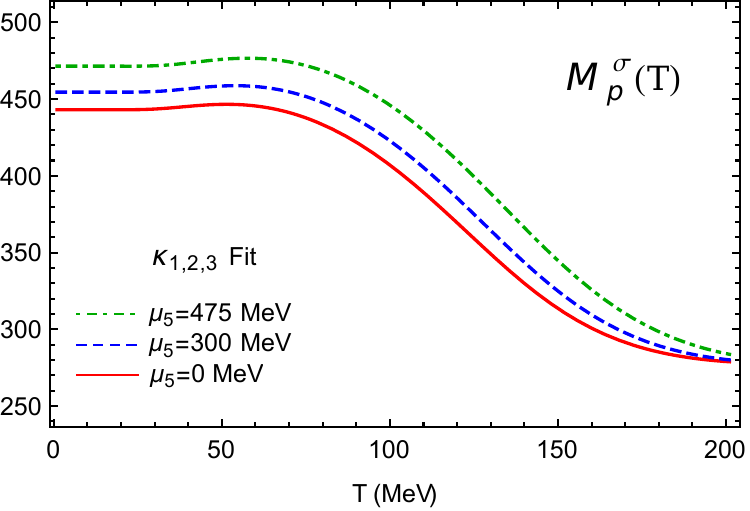}\hspace{0.5cm}
\includegraphics[width=0.45\textwidth]{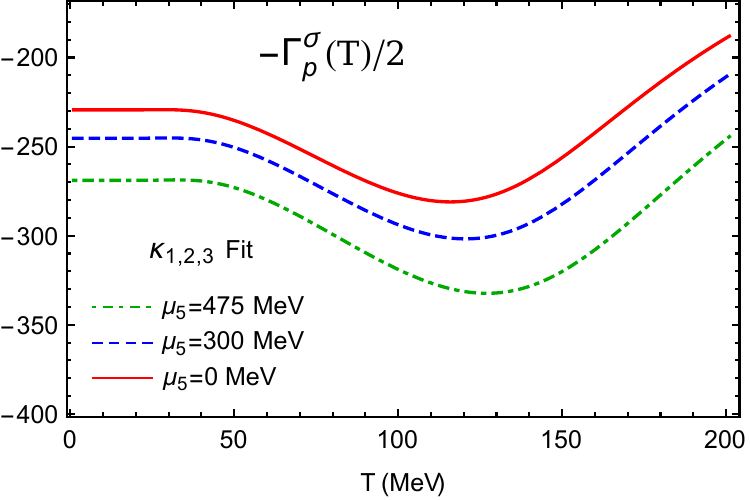}\\
\includegraphics[width=0.45\textwidth]{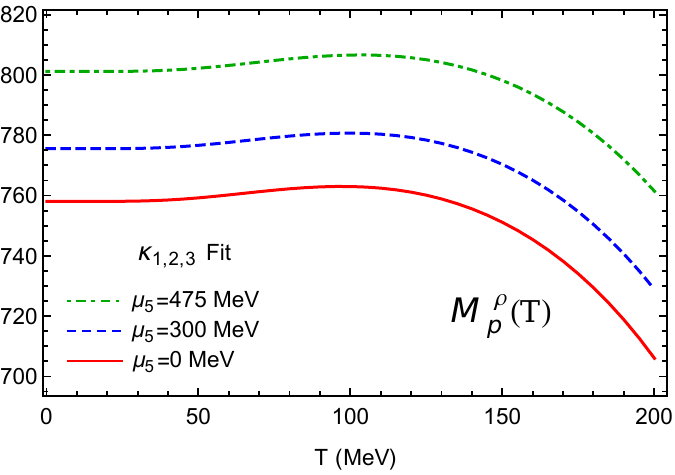}\hspace{0.5cm}
\includegraphics[width=0.45\textwidth]{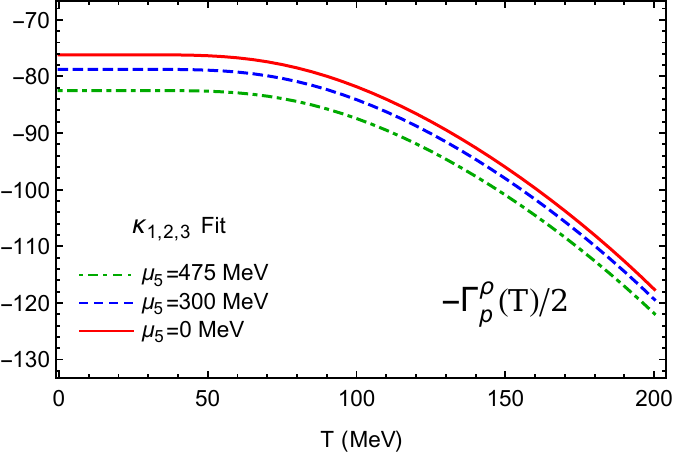}
    \caption{Temperature evolution of the mass and width of the $f_{0}(500)$ and $\rho(770)$ resonances for $\mu_{5}=0,\,300$ and $475$ MeV.}
  \label{fig:MassandwidthsigmaTev}
\end{figure}

The results for the combined $T$ and $\mu_5$ corrections for the pole parameters are showed in Fig.\ref{fig:MassandwidthsigmaTev}.  Our conclusion  is that both the mass and width pole parameters $M_p,\Gamma_p$ increase with $\mu_5$ for both channels and for all temperatures.

For the $f_0(500)$ case, the effect is numerically more significant for $\Gamma_p$, while for $M_p$ the $\mu_5$ increase tends to vanish as the temperature approaches the transition region. The latter can be understood as a chiral restoring behaviour. Namely, $M_p(T)$ is expected to decrease rapidly with $T$ driven by chiral symmetry restoration, reaching the two-pion threshold while the $f_0(500)$ scalar mass $M_S^2$, where both $M_p$ and $\Gamma_p$ enter, tends to become degenerate with the pion mass
\cite{Nicola:2020smo}. 
Thus, the increase of $M_p^\sigma$ with $\mu_5$ at any $T$ is consistent with the increase of $T_c (\mu_5)$, while the curves converging as $T$ increases indicates that the dropping effect of $M_p^\sigma (T)$ towards threshold driven by chiral restoration  is stronger than the $\mu_5$ increase.  Such behaviour is almost unaffected by the $\mu_5$ corrections to the pion mass, which, with the values of $\kappa_i$ in Table  \ref{tab:kappa} are of the order of a few percent, both for the pole and screening pion masses derived from the dispersion relation \cite{Espriu:2020dge}.

As for the $\rho$ channel, the dominant $\mu_5$ effect is the increase of the mass, while the width increase is softer,  contrary to the $T$ effect, as seen in Fig.\ref{fig:MassandwidthsigmaTev}. Regarding the  connection with the dilepton spectrum,  the combined effect of $T$ and $\mu_5$ corrections near the transition temperature   would be then a displacement (mass increase) and widening of the dilepton yield around the $\rho$ mass region. This is qualitively in agreement with the analysis  in   \cite{Chaudhuri:2022rwo} within the NJL model, where for vanishing three-momenta of the dilepton pair, which corresponds to the case of the $\rho$ at rest with the thermal bath that we are considering here, for $\mu_5\neq 0$ the unitarity cut contribution to dileptons is displaced to a higher invariant mass than for $\mu_5=0$.

\begin{figure}[h!]
    \centering
\includegraphics[width=0.6\textwidth]{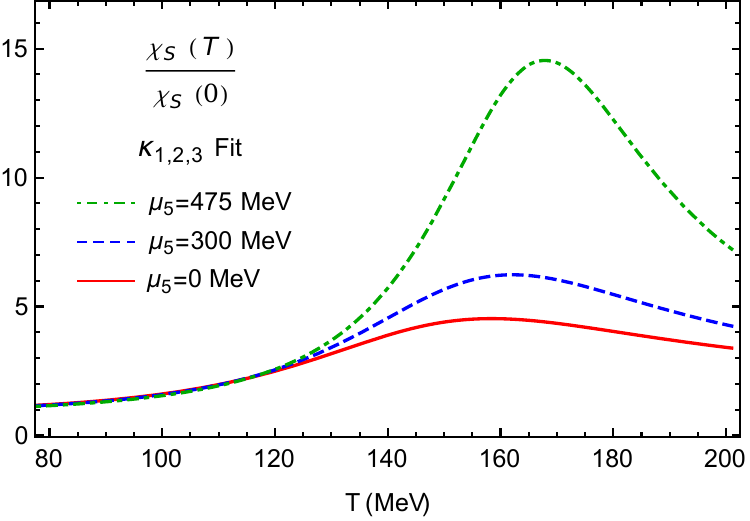}
    \caption{Scalar susceptibility for different values of $\mu_{5}$.}
  \label{fig:susmu5}
\end{figure}

Finally, in Fig.\ref{fig:susmu5} we plot the $f_0(500)$ saturated scalar susceptibility temperature dependence for $\mu_5=0,300, 475$ MeV. Apart from the shift of the peak position corresponding to the increasing of $T_c$, $\chi_S$ is significantly larger with increasing $\mu_5$, notably around $T_c$.  This qualitative behaviour is confirmed by the lattice analysis in \cite{Braguta:2015zta} although the higher pion mass used in those simulations does not allow for a direct comparison with the result in Fig.\ref{fig:susmu5}.

\section{Conclusions}
\label{sec:conc}

We have calculated  $\pi\pi$ scattering to one loop in Chiral Perturbation Theory at nonzero temperature and nonzero chiral imbalance chemical potential $\mu_5$. The ChPT amplitude has been unitarized within the inverse amplitude method, which allows to generate dynamically the light resonances $f_0(500)/\sigma$ and $\rho(770)$ and study the modification of their spectral properties with $T$ and $\mu_5$. 

From the $f_0(500)$ pole, using the saturation approach followed in previous works, we have calculated the scalar susceptibility $\chi_S (T,\mu_5)$.  For the $\mu_5$ range considered in this work, $\chi_S$ develops a peak which signals chiral symmetry restoration at $T_c(\mu_5)$, consistently with $\mu_5=0$ lattice determinations. 

Firstly, our present calculation of the ratio $T_c(\mu_5)/T_c(0)$  from $\chi_S$ allows  to improve the determination of the low-energy constants $\kappa_i$ of the $\mu_5\neq 0$ lagrangian. That ratio is actually pretty insensitive to the pion mass, which makes it a suitable quantity to compare with our theoretical predictions. Actually, in \cite{Espriu:2020dge}, the same ratio was used to fit the $\kappa_i$ combinations appearing in the quark condensate. Since the combinations appearing in $\chi_S$, coming from pion scattering, are different from those in the quark condensate, we have been actually able to reduce the uncertainties by fitting lattice points for $T_c(\mu_5)/T_c(0)$ and the topological susceptibility. An important difference of the $T_c(\mu_5)$ determination from our present analysis with respect to that from the quark condensate in  \cite{Espriu:2020dge} is that the chiral limit expressions alone are not enough to fit accurately the lattice results, since the  behaviour of the susceptibility in that limit is qualitatively very different from the massive case, corresponding to a divergence near $T_c$. 

Secondly, for the $\kappa_i$ values obtained in our main fit, both the critical temperature and the scalar susceptibility increase with $\mu_5$, in agreement with the lattice results and consistently with the analysis in \cite{Espriu:2020dge}.

Finally, our results for the phase shifts for those $\kappa_i$ show a reduction with $\mu_5$ for the three channels $IJ=00,11,20$, while the resonance pole parameters $M_p$, $\Gamma_p$ increase with $\mu_5$ for all temperatures, showing a behaviour compatible with chiral restoration in the $IJ=00$ case and with previous analysis on dilepton production in the $IJ=11$ one. 

In summary, our present work advances on the knowledge of the properties of light hadron matter within a chirally asymmetric environment, providing useful results regarding the Physics of locally $P$-breaking QCD in Heavy-Ion collisions.

\begin{acknowledgments}
 Work partially supported by research contracts  PID2019-106080GB-C21 and PID2022-136510NB-C31 (spanish ``Ministerio de Ciencia e Innovaci\'on") anb by  the European Union Horizon 2020 research and innovation program under grant agreement No 824093. A. V-R acknowledges support from  a fellowship of the UCM predoctoral program.
 \end{acknowledgments}

\end{document}